**Protected valley states and generation of valley- and spin-polarized current in monolayer $MA_2Z_4$**

Jiaren Yuan,[1] Qingyuan Wei,[1] Minglei Sun,[2] Xiaohong Yan,[1,*] Yongqing Cai,[3,†] Lei Shen (沈雷),[4,‡] and Udo Schwingenschlögl[2]

[1]*School of Physics and Electronic Engineering, and School of Material Science and Engineering, Jiangsu University, Zhenjiang 212013, China*
[2]*Physical Science and Engineering Division (PSE), King Abdullah University of Science and Technology (KAUST), Thuwal 23955-6900, Saudi Arabia*
[3]*Joint Key Laboratory of the Ministry of Education, Institute of Applied Physics and Materials Engineering, University of Macau, Macau 999078, China*
[4]*Department of Mechanical Engineering and Engineering Science, National University of Singapore, 9 Engineering Drive 1, Singapore 117542*

The optical selection rules obeyed by two-dimensional materials with spin-valley coupling enable the selective excitation of carriers. We show that six members of the monolayer $MA_2Z_4$ (M = Mo and W; A = C, Si, and Ge; Z = N, P, and As) family are direct band-gap semiconductors with protected valley states and that circularly polarized infrared light can induce valley-selective inter-band transitions. Our optovalleytronic device demonstrates a close to 100% valley- and spin-polarized current under in-plane bias and circularly polarized infrared light, which can be exploited to encode, process, and store information.

[*] yanxh@ujs.edu.cn

[†]yongqingcai@um.edu.mo

[‡]shenlei@nus.edu.sg

## I. INTRODUCTION

In some two-dimensional (2D) materials the electrons have a valley degree of freedom besides the charge and spin degrees of freedom due to the appearance of inequivalent valleys (band extrema with equal energy located at different *k*-points in the Brillouin zone) [1,2]. In the absence of inversion symmetry the spin-orbit coupling (SOC) can result in valley-dependence of the Berry curvature, orbital magnetic moment, and optical circular dichroism [3,4]. Generation of a valley-polarized current is a prerequisite for utilizing the valley degree of freedom as information carrier in valleytronic devices [5]. Given the intriguing fundamental physics and potential applications in electronics and optoelectronics, including Hall devices [6] and photoelectric detection [7], exploring materials with a valley degree of freedom and generating valley-polarized current are important research directions.

The valley properties were investigated theoretically and experimentally for 2D materials such as graphene [8,9], SnSe [10,11], $MnPSe_3$ [12,13], and transition-metal dichalcogenides (TMDCs) [2,14]. In particular, TMDCs provide an excellent platform for the study of spin-valley coupling and valley polarization [14,15]. Different from their bulk phases, monolayer TMDCs, such as $MoS_2$ and $MoSe_2$, have a direct band gap with conduction band minima (CBMs) and valence band maxima (VBMs) located at both the inequivalent *K* and *K*' points [2]. The spin and valley degrees of freedom are coupled by the time-reversal symmetry, enabling their control [14,15]. Valley polarization can be realized by breaking the time-reversal symmetry by an external magnetic field [16], proximity effects [17], magnetic doping [18], and circularly polarized light [19,20]. For example, a large valley polarization was achieved in monolayer $MoS_2$ and $WS_2$ by circularly polarized light due to valley-dependence of the optical selection rules [3,4].

The intercalated 2D semiconductor $MoSi_2N_4$ and $WSi_2N_4$ recently were synthesized by chemical vapor deposition [21] and were found to exhibit rich electronic properties [21-26]. The spin-valley coupling and valley polarization of magnetically doped monolayer $MoSi_2N_4$, $MoSi_2P_4$, and $MoSi_2As_4$ were studied in Refs. [18,23], while the properties of the other members of the monolayer $MA_2Z_4$ (M = Mo and W; A = C, Si, and Ge; Z = N, P, and As) family yet have to be investigated. In addition, there exist no insights into the behavior of the valley-polarized current for these systems. Using first-principles calculations, we thus systematically investigate the 18 monolayers. We obtain for 6 of them ($MoSi_2P_4$, $MoSi_2As_4$, $WSi_2P_4$, $WSi_2As_4$, $WGe_2P_4$, and $WGe_2As_4$) a direct band gap of 0.15-0.62 eV with band extrema located at both the inequivalent *K* and *K'* points. All show strong spin-valley coupling. We propose a device in which a valley- and spin-polarized current can be generated by circularly polarized light. Our results demonstrate that monolayer $MA_2Z_4$ offers new opportunities for valley-based electronic and optoelectronic applications.

## II. METHODOLOGY

Electronic structure calculations are carried out by the Vienna ab initio Simulation Package [27], utilizing the projector augmented wave method [28] and a cutoff energy of 450 eV. The Perdew-Burke-Ernzerhof (PBE) [29] and Heyd-Scuseria-Ernzerhof (HSE06) [30] approaches are used for the exchange correlation functional. A 19 × 19 × 1 k-grid is employed. The convergence criterion of the total energy is set to $10^{-6}$ eV and that of the force to 0.005 eV/Å. Spin-orbit coupling is taken into account in all calculations. The Berry curvature is calculated by the Wannier90 code [31]. The photocurrent is calculated by the quantum transport package Nanodcal[32-35] based on the Hamiltonian $H = H_{\mathrm{el}} + H_{\mathrm{el\text{-}ph}}$. The electronic contribution $H_{\mathrm{el}}$ is calculated using the first-principles and Keldysh non-equilibrium Green's function methods. The electron-photon interaction $H_{\mathrm{el\text{-}ph}}$ is calculated perturbatively using the first Born approximation, $H_{\mathrm{el-ph}} = \frac{e}{m}\boldsymbol{A}\cdot\boldsymbol{p}$, where $\boldsymbol{A}$ is the electromagnetic vector potential and $\boldsymbol{p}$ is the momentum of the electron. For circularly polarized light, we have $\boldsymbol{A} = (\frac{\hbar\sqrt{\tilde{\mu}_r\tilde{\epsilon}_r}}{2N\omega\tilde{\epsilon}c}I_\omega)^{1/2}(\boldsymbol{e}_p\mathrm{b}e^{-i\omega t} - \boldsymbol{e}_p^{*}b^{\dagger}e^{i\omega t})$, where $\omega$ is the frequency, $I_\omega$ is the photon flux, $c$ is the speed of light, $\tilde{\mu}_r$ is the relative magnetic susceptibility, $\tilde{\epsilon}_r$ is the relative dielectric constant, $\tilde{\epsilon}$ is the dielectric constant, $N$ is the number of photons, $b$ and $b^{\dagger}$ are the annihilation and creation operators, respectively, and $\boldsymbol{e}_p = \frac{1}{\sqrt{2}}(1, \pm \mathrm{i}, 0)$ for left-handed/right-handed circularly polarized light. The current is given by $I_{\alpha,\tau,s} =$

$\frac{e}{2\pi\hbar}\int \mathrm{Tr}\{i\Gamma_\alpha(E,k)[(1-f_\alpha)G^{<}_{ph}+f_\alpha G^{>}_{ph}]\}dE$, where $\Gamma_\alpha(E,k)$ is the line-width, $E$ is the energy, $k$ is the wave vector, $f_\alpha$ is the Fermi function, $\alpha$ denotes the left/right electrode, $\tau$ denotes the $K$/$K$' valley, $s$ denotes spin up/down, and $G^{<}_{ph}/G^{>}_{ph}$ is the greater/lesser Green's function of the electron-photon interaction.

## III. RESULTS AND DISCUSSION

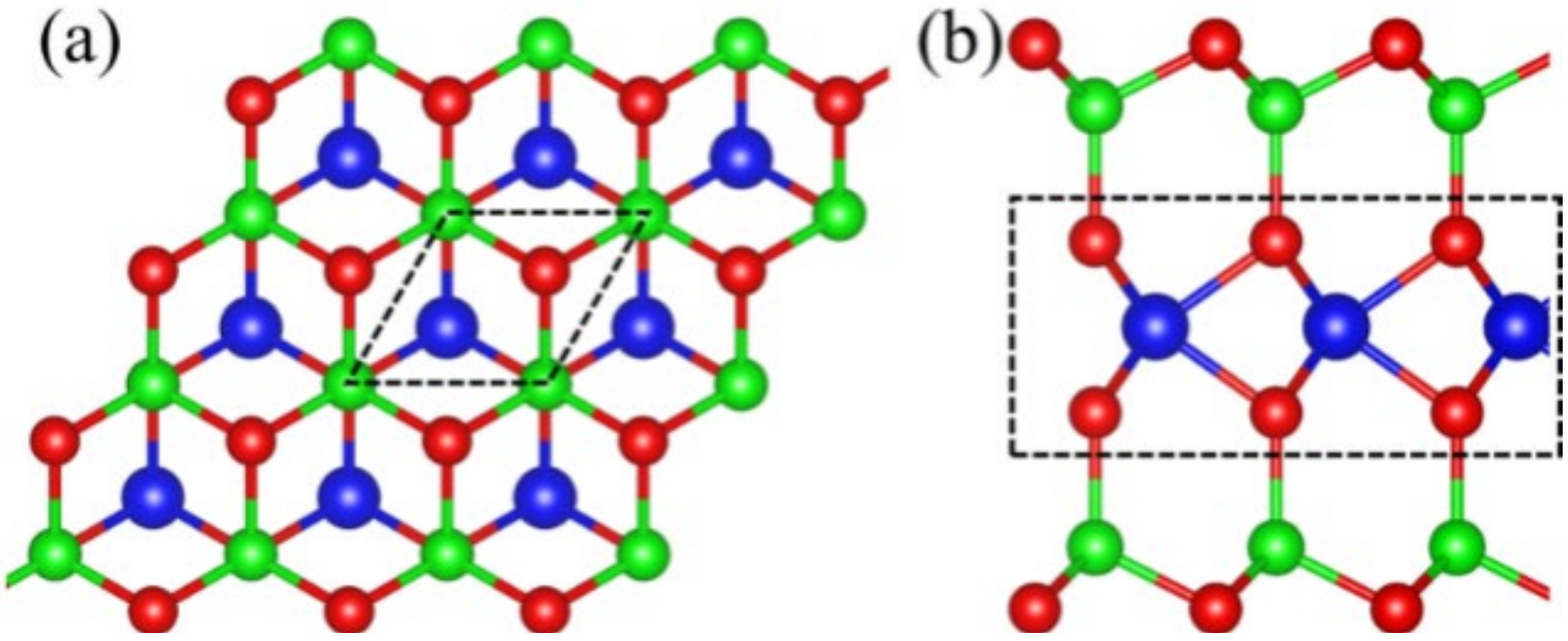

**FIG. 1** (a) Top and (b) side views of the relaxed structure of $MA_2Z_4$. The blue, red, and green balls represent the M, A, and Z atoms, respectively. The unit cell is marked by dashed lines in (a). The structure can be regarded as a $MZ_2$ triple-layer (resembling 1H-phase $MoS_2$; dashed rectangle in (b)) encapsulated by buckled AZ layers.

Figure 1 shows that monolayer $MA_2Z_4$ (M = Mo and W; A = C, Si, and Ge, Z = N, P, and As) has a hexagonal lattice with $D_{3h}$ point group, that is, without inversion symmetry, and consists of seven atomic layers of the order Z-A-Z-M-Z-A-Z. It can be regarded as a 1H-phase $MZ_2$ triple-layer encapsulated by buckled AZ layers. The optimized lattice constants listed in Table 1 increase with the atomic radii of M, A, and Z.

**Table 1** Optimized lattice constants and PBE band gaps (HSE06 band gaps in brackets).

| $MA_2Z_4$ | $a$ (Å) | $E_{gap}$ (eV) | Type of band gap |
|---|---|---|---|
| $MoC_2N_4$ | 2.621 | 1.75 | Indirect |
| $MoC_2P_4$ | 3.076 | 0.14 | Direct at $M$ |
| $MoC_2As_4$ | 3.257 | --- | --- |
| $MoSi_2N_4$ | 2.911 | 1.78 | Indirect |
| $MoSi_2P_4$ | 3.469 | 0.62 (0.86) | Direct at *K* |
| $MoSi_2As_4$ | 3.617 | 0.51 (0.71) | Direct at *K* |
| $MoGe_2N_4$ | 3.036 | 0.90 | Indirect |
| $MoGe_2P_4$ | 3.548 | 0.52 | Indirect |
| $MoGe_2As_4$ | 3.691 | 0.43 | Indirect |
| $WC_2N_4$ | 2.638 | 1.66 | Indirect |
| $WC_2P_4$ | 3.088 | --- | --- |
| $WC_2As_4$ | 3.264 | --- | --- |
| $WSi_2N_4$ | 2.913 | 2.05 | Indirect |
| $WSi_2P_4$ | 3.475 | 0.29 (0.41) | Direct at *K* |
| $WSi_2As_4$ | 3.622 | 0.21 (0.22) | Direct at *K* |
| $WGe_2N_4$ | 3.034 | 1.13 | Indirect |
| $WGe_2P_4$ | 3.552 | 0.23 (0.29) | Direct at *K* |
| $WGe_2As_4$ | 3.697 | 0.15 (0.14) | Direct at *K* |

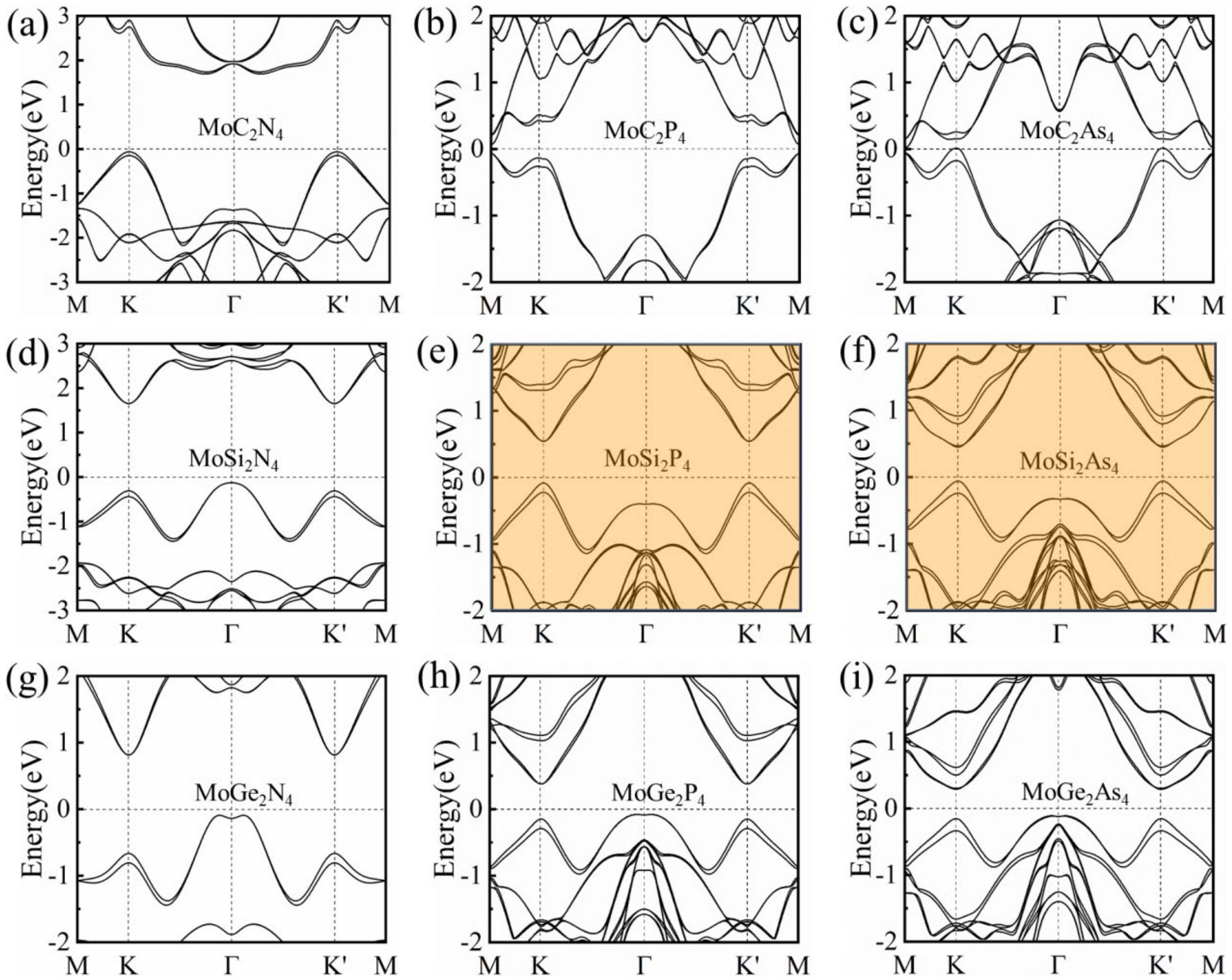

**FIG. 2** (a)-(i) Band structures of monolayer $MoA_2Z_4$. The ones with the direct band gap are highlighted in orange.

The band structures of monolayer $MA_2Z_4$ are shown in Figs. 2 and 3, and the sizes and types of the band gaps are listed in Table 1. The results reflect rich electronic structures: $MoC_2N_4$, $MoSi_2N_4$, $MoGe_2N_4$, $MoGe_2P_4$, $MoGe_2As_4$, $WC_2N_4$, $WSi_2N_4$, and $WGe_2N_4$ are indirect band gap semiconductors, $MoC_2As_4$, $WC_2P_4$, and $WC_2As_4$ are metals, and $MoC_2P_4$ is a direct band gap semiconductor (better light absorption than an indirect band gap semiconductor) with the CBM and VBM located at the *M* point. Most interestingly, $MoSi_2P_4$, $MoSi_2As_4$, $WSi_2P_4$, $WSi_2As_4$, $WGe_2P_4$, and $WGe_2As_4$ turn out to be direct band gap semiconductors with CBMs and VBMs located at both the inequivalent *K* and *K*' points (corners of the hexagonal Brillouin zone). Therefore, these 6 materials compete with the monolayer TMDCs as platform for light-controlled valleytronics and are studied in the following in more detail.

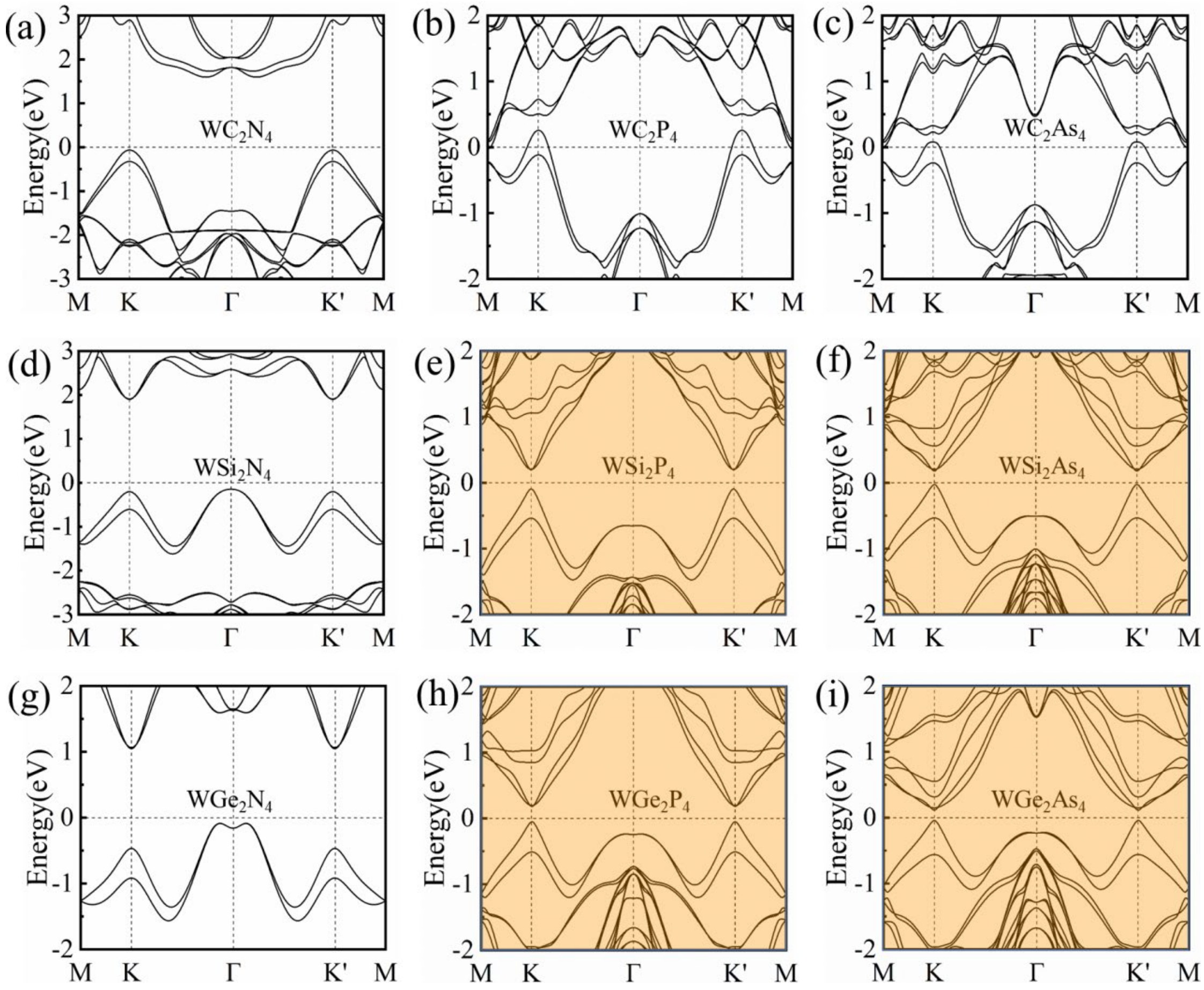

**FIG. 3** (a)-(i) Band structures of monolayer $WA_2X_4$. The ones with the direct band gap are highlighted in orange.

Spin-projected (onto the z-axis) band structures are shown in Fig. 4. Red (blue) color represents the dominance of the spin-up (spin-down) component. The PBE band gap ranges from 0.15 to 0.62 eV, as listed in Table 1. Comparison to HSE06 band structures, as displayed in Fig. S1, shows that the main band features are the same. The HSE06 band gaps, as listed in Table 1, deviate only slightly from the PBE band gaps, suggesting reliability of the PBE results. The SOC gives rise to a remarkable spin splitting near the VBM (139 to 500 meV) and a negligible spin splitting near the CBM. As expected, the W-based compounds show larger spin splitting than the Mo-based compounds. Near the VBM the upper (lower) band is dominated by the spin-up (spin-down) component at the *K* point and spin-down (spin-up) component at the *K*' point, indicating spin-valley coupling. As the band structures in Fig. 4 are very similar, we analyze the valley properties of $MoSi_2P_4$ as a representative example.

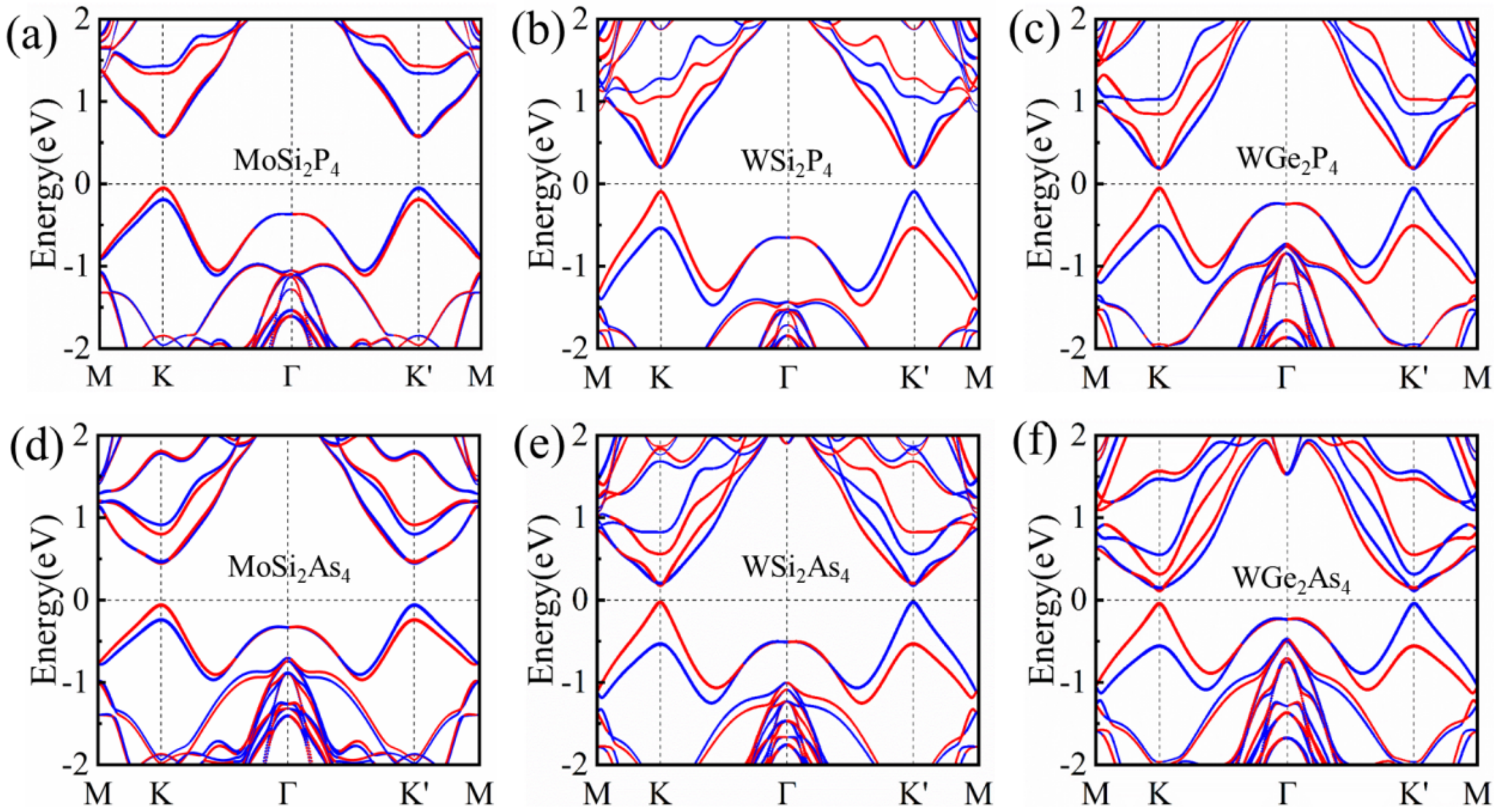

**FIG. 4** Band structures of monolayer $MoSi_2P_4$, $WSi_2P_4$, $WGe_2P_4$, $MoSi_2As_4$, $WSi_2As_4$, and $WGe_2As_4$. Red (blue) color represents dominance of the spin-up (spin-down) component.

Figure 5(a) shows schematically the valleys at the *K* and *K*' points. Time-reversal symmetry leads to opposite spin polarization near the VBM at the *K* and *K*' points. The charge densities at the VBM and CBM in Fig. 5(b) show that the valleys originate mainly from the $MoP_2$ layer. This conjecture is confirmed by similarity to the valleys in the band structure of 1H-phase $MoP_2$ (without the buckled SiP layers), as shown in Fig. 5(c). However, there are interfering states at other $k$ points in the energy range of the valleys in Fig. 5(c). Hence, the buckled SiP layers of $MoSi_2P_4$ not only stabilize the structure but also protect the valley states from interfering P $p_z$ and Mo $d_{3z^2-r^2}$ states (see Fig. S2). The partial densities of states displayed in Fig. 5(d) demonstrate that the protected valley states are dominated by the Mo d orbitals, which split into $a_1$ ($d_{3z^2-r^2}$), $e_1$ ($d_{x^2-y^2}$, $d_{xy}$), and $e_2$ ($d_{xz}$, $d_{yz}$) groups due to the trigonal crystal field of the $MoP_2$ layer (see Fig. S3). Specifically, the CBM is mainly due to the Mo $a_1$ states and the VBM is mainly due to the Mo $e_1$ states, as shown in the orbital-projected band structure in Fig. 5(e).

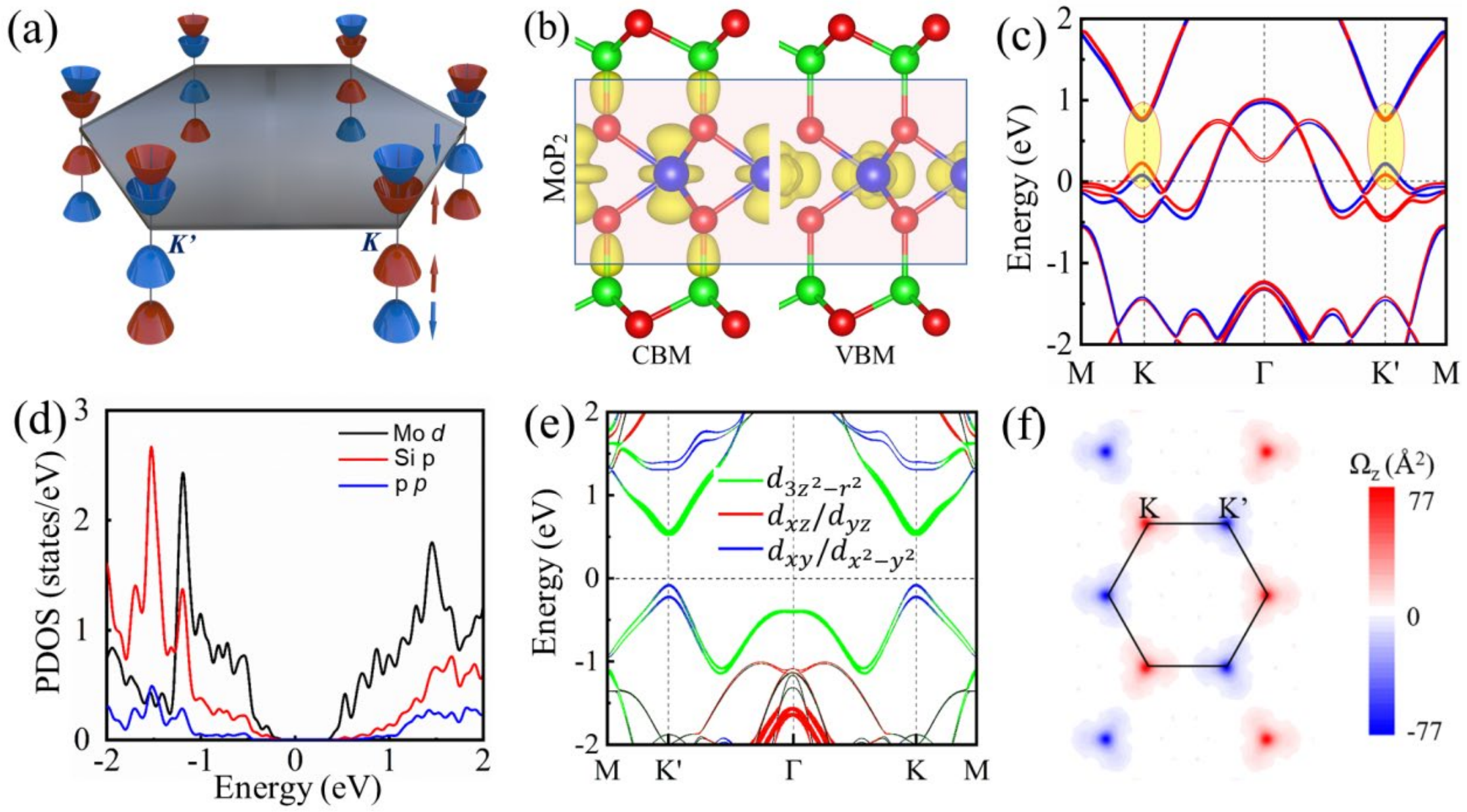

**FIG. 5** (a) Schematic of the valleys near the $K$ and $K'$ points, (b) charge densities at the VBM and CBM of $MoSi_2P_4$, (c) band structure of 1H-phase $MoP_2$, (d) partial densities of states of $MoSi_2P_4$, (e) orbital-projected band structure of $MoSi_2P_4$, and (f) Berry curvature of $MoSi_2P_4$.

Consequently, we define the basis functions

$$|\varphi_c\rangle = |d_{3z^2-r^2}\rangle; \quad |\varphi_v^\tau\rangle = \frac{1}{\sqrt{2}}\left(|d_{x^2-y^2}\rangle + i\tau|d_{xy}\rangle\right), \tag{1}$$

where c (v) denotes the conduction (valence) band and $\tau = +1$ (–1) denotes the $K$ ($K'$) valley, to express the Hamiltonian of the two-band effective $\boldsymbol{k} \cdot \boldsymbol{p}$ model without SOC as

$$H_0^\tau = at(\tau k_x \hat{\sigma}_x + k_y \hat{\sigma}_y) + \frac{\Delta}{2}\hat{\sigma}_z, \tag{2}$$

where $a$ denotes the lattice parameter, $t$ denotes the effective hopping integral, $\hat{\sigma}_{x/y/z}$ denotes the Pauli matrices, and $\Delta$ denotes the band gap. Due to the common symmetry, the same model applies to the 1H-phase TMDCs [2], implying that the low-energy band structures are equivalent. When the SOC is taken into account, we have

$$H_0^\tau = at(\tau k_x \hat{\sigma}_x + k_y \hat{\sigma}_y) + \frac{\Delta}{2}\hat{\sigma}_z - \lambda\tau\frac{\hat{\sigma}_z - 1}{2}\hat{s}_z, \tag{3}$$

where $2\lambda$ is the spin-splitting at the VBM (induced by the SOC) and $\hat{s}_z$ is the Pauli operator. The effective parameters of the $\boldsymbol{k} \cdot \boldsymbol{p}$ model, see Table 2, are extracted from the first-principles band structures. We find that $t$ is smaller for $MoA_2Z_4$ than $WA_2Z_4$, while $\Delta$ shows the opposite trend with smaller values than reported for the 1H-phase TMDCs. The corresponding HSE06 values are larger but follow the same trends as the PBE values (Table 2).

**Table 2**. Effective hopping integral, band gap, and spin-splitting extracted from the first-principles band structures.

| Material | Method | $t$ (eV Å) | $\Delta$ (eV) | $2\lambda$ (eV) |
|---|---|---|---|---|
| $MoSi_2As_4$ | PBE | 2.44 | 0.61 | 0.18 |
| | HSE06 | 3.18 | 0.86 | 0.24 |
| $MoSi_2P_4$ | PBE | 2.36 | 0.70 | 0. 14 |
| | HSE06 | 3.80 | 0.98 | 0.22 |
| $WGe_2As_4$ | PBE | 2.84 | 0.45 | 0.50 |
| | HSE06 | 3.70 | 0.65 | 0.73 |
| $WGe_2P_4$ | PBE | 3.16 | 0.48 | 0.45 |
| | HSE06 | 4.45 | 0.68 | 0.61 |
| $WSi_2As_4$ | PBE | 3.15 | 0.49 | 0.50 |
| | HSE06 | 4.18 | 0.70 | 0.70 |
| $WSi_2P_4$ | PBE | 3.77 | 0.53 | 0.44 |
| | HSE06 | 4.78 | 0.76 | 0.59 |

To analyze the valley properties, we evaluate the out-of-plane Berry curvature

$$\Omega_z(k) = -\sum_n \sum_{n\neq n'} f \frac{2\mathrm{Im}\langle\psi_{nk}|v_x|\psi_{n'k}\rangle\langle\psi_{n'k}|v_y|\psi_{nk}\rangle}{(E_n - E_{n'})^2}, \qquad (4)$$

where $f$ is the Fermi-Dirac distribution function, $v_x$ $(v_y)$ is the velocity operator for the $x$ ($y$) direction, and $\psi_{nk}$ is the Bloch function with eigenvalue $E_n$. The Berry curvature summed over 46 bands is shown in Fig. 5(f). We obtain the same absolute values but with opposite sign at the $K$ and $K$' valleys. No valley-polarized current is generated under time-reversal symmetry due to equal contributions of the $K$ and $K$' valleys. Application of circularly polarized light breaks the time-reversal symmetry, with the circular polarization $\eta$ being proportional to the Berry curvature[36]. As the inter-band transitions obey different optical selection rules at the $K$ and $K'$ valleys, the electrons couple with left and right circularly polarized light, respectively. Thus, one can optically induce valley polarization and generate a valley-polarized current in monolayer $MA_2Z_4$ by excitation with circularly polarized light.

Finally, we propose an optovalleytronic device based on monolayer $MA_2Z_4$ (Fig. 6(a)) that can generate a valley-polarized current under in-plane bias. Circularly polarized (monochromatic) light illuminates the channel between (semi-infinite) source and drain electrodes. Focusing again on $MoSi_2P_4$ as a representative example, we set the photon energy equal to the band gap, namely $\hbar\omega$ = 0.62 eV, implying that only electrons located at the $K$ and $K$' points can be excited. In the rectangular transport setup, the $K$ and $K$' points are folded onto the points (–1/3, 0, 0) and (1/3, 0, 0). The drain-source voltage $V_{ds}$ generates a valley-polarized current, see the schematic in Fig. 6(b). When circularly polarized light illuminates the channel and breaks the time-reversal symmetry, the electrons at the $K$ and $K$' valleys absorb photons (and are excited from the VBM to the CBM) differently, which induces a population imbalance between the $K$ and $K$' valleys. Under an in-plane bias the excited electrons flow into the drain. At the same time, holes are left behind in the channel and are filled by electrons from the source, resulting in an overall current from the source to the drain.

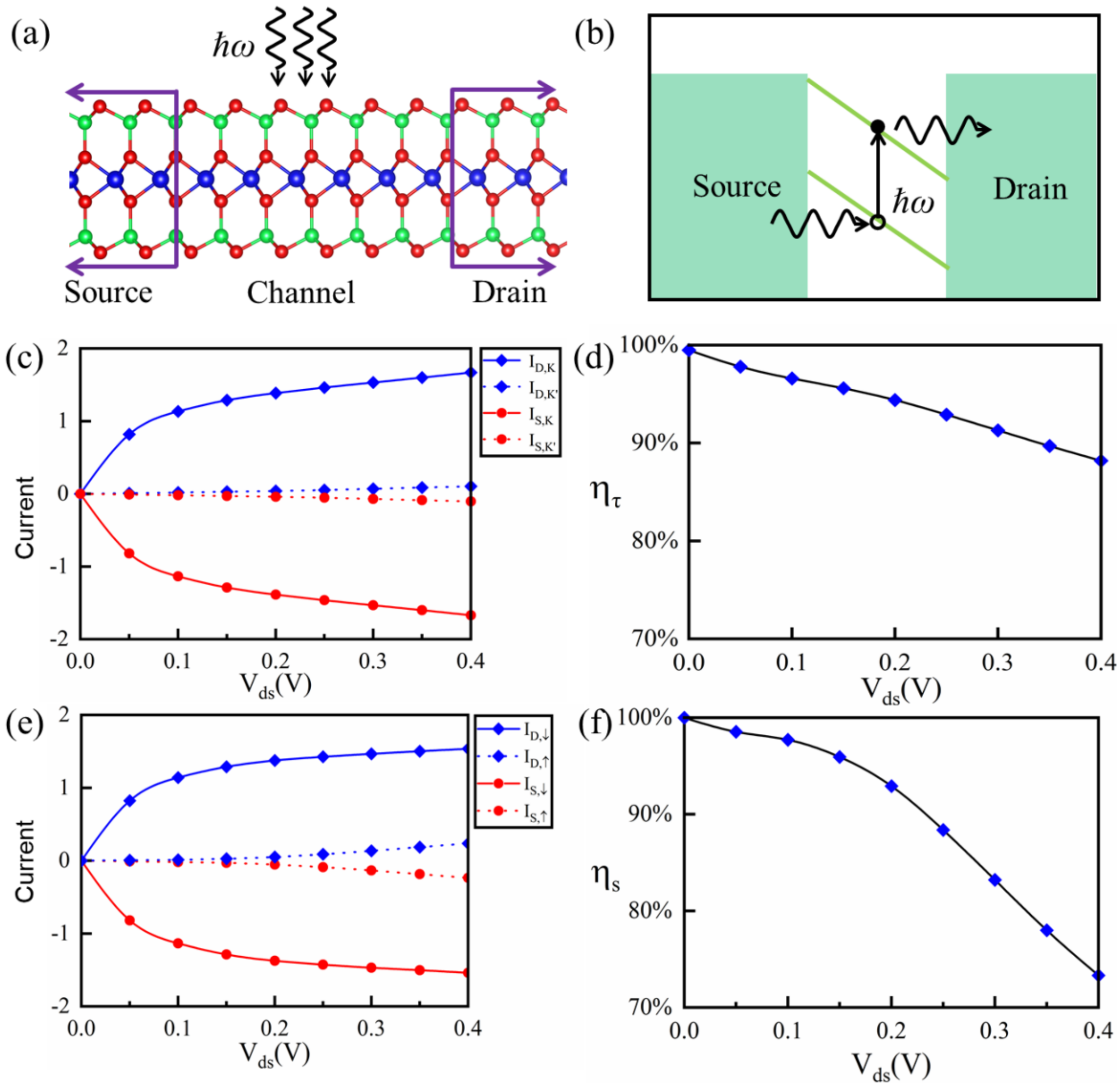

**FIG. 6** (a)-(b) Schematic of the monolayer $MA_2Z_4$ device and its generation of valley-polarized current. (c)-(f) valley currents, valley polarization, spin currents, and spin polarization of the monolayer $MoSi_2P_4$ device.

The current generated by right-handed circularly polarized light is investigated for $V_{ds} < \Delta/e$, that is, no direct current is generated. The source and drain valley currents consist of spin-up and spin-down contributions, $I_{S/D,\tau} = I_{S/D,\tau,\uparrow} + I_{S/D,\tau,\downarrow}$. According to Fig. 6(c), the $K$ valley currents are much larger than the $K$' valley currents, because the right-handed circularly polarized light excites electrons rather at the $K$ than at the $K$' valley due to the optical selection rules. Thus, a valley-polarized current $I^{\tau}_{S/D}$ = $I_{S/D,K} - I_{S/D,K'}$ is generated between the source and drain. This current is close to $I_{S/D,K}$, since $I_{S/D,K'}$ is small, see Fig. 6(d). Accordingly, a tremendous valley polarization $\eta_\tau$ =

$\frac{I_{S/D,K} - I_{S/D,K'}}{I_{S/D,K} + I_{S/D,K'}}$ is achieved, see Fig. 6(d), close to 100% at zero bias. It decreases for increasing bias, because the bias breaks the time-reversal symmetry. However, note that this effect will decay when the channel length increases [34]. As the spin degree of freedom is locked to the valley degree of freedom, see the discussion above, spin currents appear simultaneously with the valley currents. Evaluation by summation over the *K* and *K*' valleys shows that the spin currents, spin-polarized current, and spin polarization follow the same trends as their valley analogues, see Figs. 6(e) and 6(f).

## IV. CONCLUSION

Six members of the monolayer $MA_2X_4$ (M = Mo and W; A = C, Si, and Ge; Z = N, P, and As) family are found to be direct band-gap semiconductors with CBMs and VBMs located at both the inequivalent *K* and *K*' points. These valley states originate mainly from the $MZ_2$ triple-layer and are protected from interfering states due to the encapsulation by buckled AZ layers. Furthermore, the states near the VBM are subject to strong spin-valley coupling and significant spin splitting due to the absence of inversion symmetry and the presence of strong SOC. The direct band gap is smaller than in the case of monolayer TMDCs, falling into the infrared spectral range. Optical pumping by circularly polarized infrared light thus can induce valley polarization in monolayer $MA_2Z_4$ and enables the generation of a valley- and spin-polarized current under an in-plane bias. Our results demonstrate that monolayer $MA_2Z_4$ provides an alternative platform for investigating the interplay of the spin and valley degrees of freedom, pushing forward the development of quantum manipulation in valley-based electronic and optoelectronic devices.

## ACKNOWLEDGEMENTS

This work was supported by the National Natural Science Foundation of China (NSFC12004142 and NSFC12174158), China Postdoctoral Science Foundation (2020M1350), Natural Science Funds for Colleges and Universities in Jiangsu

Province (20KJB140017), Postdoctoral Research Funding Program of Jiangsu Province (2020Z131), and Singapore Ministry of Education Academic Research Fund Tier 1 Grants No. R-265-000-651-114 and No. R265-000-691-114. The research reported in this publication was supported by funding from King Abdullah University of Science and Technology (KAUST)